# Structure evolution of hcp/ccp metal oxide interfaces in solid-state reactions

C. Li[a,b]*, G. Habler[a], T. Griffiths[a], A. Rečnik[c], P. Jeřábek[d], L. C. Götze[e], C. Mangler[f], T. J. Pennycook[f,b], J. Meyer[f], & R. Abart[a]

[a] Department of Lithospheric Research, University of Vienna, Althanstrasse 14, 1090 Vienna, Austria

[b] Stuttgart Center for Electron Microscopy, Max Planck Institute for Solid State Research, Heisenbergstr. 1, 70569 Stuttgart, Germany

[c] Department for Nanostructured Materials, Jožef Stefan Institute, Jamova cesta 39, SI-1000 Ljubljana, Slovenia

[d] Institute of Petrology and Structural Geology, Faculty of Science, Charles University, Albertov 6, 12843 Prague, Czech Republic

[e] Institute of Geological Sciences, Freie Universität Berlin, Malteserstr. 74-100, 12249 Berlin, Germany

[f] Faculty of Physics, University of Vienna, Boltzmanngasse 5, 1090 Vienna, Austria

* Corresponding author. Email: chen.li@fkf.mpg.de

**Synopsis** The atomic structure of $Al_2O_3$/$MgAl_2O_4$ interfaces at different growth stages are revealed by scanning transmission electron microscopy. Partial dislocations in the *hcp*/*ccp* oxygen sublattices become increasingly dominant as the growth proceeds, suggesting a dislocation glide mechanism in the late growth stage.

**Abstract** The structure of crystalline interfaces plays an important role in solid-state reactions. The $Al_2O_3$/$MgAl_2O_4$/$MgO$ system provides an ideal model system for investigating the mechanisms underlying the migration of interfaces during interface reaction. $MgAl_2O_4$ layers have been grown between $Al_2O_3$ and $MgO$, and the atomic structure of $Al_2O_3$/$MgAl_2O_4$ interfaces at different growth stages was characterized using aberration-corrected scanning transmission electron microscopy. The oxygen sublattice transforms from hexagonal close-packed (*hcp*) stacking in $Al_2O_3$ to cubic close-packed (*ccp*) stacking in $MgAl_2O_4$. Partial dislocations associated with steps are observed at the interface. At the reaction-controlled early growth stages, such partial dislocations coexist with the edge dislocations. However, at the diffusion-controlled late growth stages, such partial dislocations are dominant. The observed structures indicate that progression of the $Al_2O_3$/$MgAl_2O_4$ interface into $Al_2O_3$ is accomplished by the glide of partial dislocations accompanied by the exchange of $Al^{3+}$ and $Mg^{2+}$ cations. The interface migration may be envisaged as a plane-by-plane zipper-like motion, which repeats along the interface facilitating its propagation. $MgAl_2O_4$ grains can adopt two crystallographic orientations with a twinning orientation relationship, and grow by dislocations gliding in opposite directions. Where the oppositely propagating partial dislocations and interface steps meet, interlinked twin boundaries and incoherent Σ3 grain boundaries form. The newly grown $MgAl_2O_4$ grains compete with each other, leading to a growth-selection and successive coarsening of the $MgAl_2O_4$ grains. This understanding could help to interpret the interface reaction or phase transformation of a wide range of materials that exhibit a similar *hcp/ccp* transition.

## 1. Introduction

When two solids with different compositions are put into contact at high temperature, they may react and produce an intermediate solid phase at their interface. Such solid-state reactions happen during mineral evolution in nature (Gaidies *et al.*, 2017), and have also been applied to synthesize functional materials such as oxides (Wu *et al.*, 1987; Zhao *et al.*, 2017) and alloys (Schwarz & Johnson, 1983). During such a solid-state reaction, two processes must occur. One is *interface reaction*, including all processes that proceed localized at the interface, where the reactant phases are replaced by the product phase. Another one is *elemental diffusion*, which is required to transfer elements across the growing product layer (Philibert, 1991; Gaidies *et al.*, 2017). The entire process including *interface reaction* and *elemental diffusion* is called *reactive diffusion*. The layer of the newly grown phase (*i.e.* the interlayer) can be only a few nanometers thick in semiconductor thin films (Li *et al.*, 2014) and nanomaterials (Fan *et al.*, 2007; 2006), but can also grow to micrometer-scale such as in oxide ceramics (Hesse, 1987; Keller *et al.*, 2010; Götze *et al.*, 2009) and natural minerals (Vernon, 2004; Pitra *et al.*, 2010; Keller *et al.*, 2008; 2006). In mineralogy, this interlayer is often referred to as a reaction rim, or a corona when the product phase completely encloses one of the reactant minerals.

In general, when the product layer is very thin, material transport across the layer is fast and the supply of chemical components required for reaction progress can be considered to be infinite. At this stage, the interface reaction limits the growth rate, and hence this early growth stage is "interface-reaction controlled" with a linear growth behaviour. When the interlayer grows thicker, the diffusion path becomes successively longer, and elemental diffusion across the interlayer becomes the rate-limiting factor. This late growth stage is thus "diffusion-controlled", and the layer thickness increases with the square root of time, which is referred to as parabolic growth behaviour (Abart & Petrishcheva, 2011). At all growth stages, the arrangement of atoms at the interfaces between the reactant and the product phases plays a key role in the transformation process. Unravelling the atomic structure of the interfaces is crucial for understanding the kinetics of solid-state reactions (Abart *et al.*, 2016). Unless the interface is perfectly coherent, misfit dislocations are needed at the interface to accommodate the mismatch between the two contacting lattices, and they might glide or climb during the interface migration. Glide is *conservative*, proceeding without the addition or removal of material to or from the interface, whereas climb is *non-conservative*, requiring the addition or removal of material (Balluffi *et al.*, 2005). Considering that in interlayer growth, the reaction kinetics evolve from interface-reaction controlled during early stages to diffusion-controlled at late growth stages, it is thus particularly important to understand the evolution of the atomic structures at the reaction fronts.

Due to the simple crystal structures of the phases involved and the feasibility of their growth in the laboratory, the spinel forming reaction, MgO (Periclase) + α-$Al_2O_3$ (Corundum) = $MgAl_2O_4$ (Spinel), is an ideal model system for studying solid-state reactions (Kotula *et al.*, 1998; Götze *et al.*, 2009; Sieber, Hesse, & Werner, 1997; Sieber, Werner *et al.*, 1997; Winterstein *et al.*, 2016). Besides, the structure of these three common oxides are representative of many other metal oxides used for functional applications (Fernández-García *et al.*, 2004; Sui & Charpentier, 2012; Li *et al.*, 2015; Cao *et al.*, 2016; Zhu *et al.*, 2013; Maiyalagan *et al.*, 2014; Zhao *et al.*, 2017). All three oxides have close-packed oxygen sublattices (O-sublattices): cubic close-packed (*ccp*) for both MgO and $MgAl_2O_4$, and pseudo-hexagonal close-packed (*hcp*) for α-$Al_2O_3$ (henceforth abbreviated as $Al_2O_3$). The *ccp* lattice is also referred to as the face-centered cubic (*fcc*) lattice, here we use *ccp* to emphasise its relation to the *hcp* variant. In ionic crystals, interlayer growth occurs by the interdiffusion of cations inside a stationary anion sublattice (Koch & Wagner, 1936). Accordingly, in $MgAl_2O_4$ formation, $Al^{3+}$ cations diffuse from $Al_2O_3$ towards MgO, whereas $Mg^{2+}$ cations diffuse from MgO towards $Al_2O_3$ (Carter, 1961), where two $Al^{3+}$ are exchanged for three $Mg^{2+}$ due to the charge balance requirements (Fig. 1).

The atomic configuration at the $MgAl_2O_4$/MgO interface is relatively straightforward as both $MgAl_2O_4$ and MgO have *ccp* O-sublattices (Li, Griffiths *et al.*, 2016). The $Al_2O_3$/$MgAl_2O_4$ interface is more complicated, because the O-sublattice must be transformed at the interface from *hcp* of $Al_2O_3$ to *ccp* of $MgAl_2O_4$. Two decades ago, transmission electron microscopy (TEM) was used to study the interface structure of this $AB_2O_4$ oxide system (Sieber, Werner *et al.*, 1997; Sieber, Hesse, & Werner, 1997; Hesse, 1987; Hesse *et al.*, 1994; Sieber *et al.*, 1996; Carter & Schmalzried, 1985; Li *et al.*, 1992; Senz *et al.*, 2001). The configuration of misfit dislocations was shown to be different in different growth stages, playing an important role during the migration of *ccp*/*ccp* interfaces. For *ccp*/*ccp* interface, migration by dislocation glide was documented as more prominent during the early growth stage, while migration by dislocation climb becomes more prominent during late growth stages (Sieber, Hesse, Werner *et al.*, 1997; Li, Griffiths *et al.*, 2016). However, the atomic structure and behavior of misfit dislocations at the *hcp*/*ccp* interface is still under debate. Theoretical modeling has demonstrated that *hcp* to *ccp* transformations can occur by the movement of a set of Shockley partial dislocations, which are associated with slips on the interfacial closed-packed plane (Bhadeshia, 2006; Carter & Schmalzried, 1985; Porter *et al.*, 2009). Surprisingly, previous research did not observe partial dislocations at either $MgAl_2O_4$/$Al_2O_3$ or $CoAl_2O_4$/$Al_2O_3$ interfaces. Instead, a small angle rotation between $MgAl_2O_4$ ($CoAl_2O_4$) and $Al_2O_3$ was considered to be the key for matching the *ccp* and *hcp* O-sublattices (Carter & Schmalzried, 1985; Hesse *et al.*, 1994; Sieber *et al.*, 1996).

The recent development of aberration correctors has improved the resolution of transmission electron microscopes from nm-scale to sub-Å scale, offering unprecedented possibilities for unraveling the detailed structure of interfaces. Revisiting the MgAl$_2$O$_4$/MgO (*ccp*/*ccp*) interface using state-of-art aberration-corrected scanning transmission electron microscopy (STEM) has indeed pushed our knowledge to a finer scale: the MgAl$_2$O$_4$/MgO interface was shown to have a 3-dimensional scalloped geometry with misfit dislocations located at the cusp positions (Li, Griffiths *et al.*, 2016). It is expected that the power of modern aberration-corrected STEM should also provide a better understanding of the atomic structure at the Al$_2$O$_3$/MgAl$_2$O$_4$ interface, and help to solve the puzzle of how the *hcp* O-sublattice of Al$_2$O$_3$ transforms into the *ccp* O-sublattice of MgAl$_2$O$_4$. In this work, MgAl$_2$O$_4$ layers with different thicknesses have been selected for microscopic study. In particular we investigated the atomic configurations of Al$_2$O$_3$/MgAl$_2$O$_4$ interfaces as well as the texture of the MgAl$_2$O$_4$ layers, formed during the early and the late growth stages as presented in result sections 3.1 and 3.2, respectively. In discussion section 4.1 we infer the dislocation glide and interface migration mechanism. In discussion section 4.2 we compare the difference of interface structure between the early and late growth stages, and interpret the evolution of the interface migration mechanism during the growth. Then we present the impact for further work in discussion section 4.3 and finally give a conclusion in section 5. Our new findings improve fundamental understanding of the mechanisms underlying the migration of *hcp*/*ccp* interfaces during a solid-state reaction. This is useful for further developing crystal growth of functional oxides as well as inferring growth stages of natural mineral reactions.

## 2. Experimental methods

### 2.1. Material synthesis targeted at different growth stages

MgAl$_2$O$_4$ layers representing early and late growth stages were produced using two different methods. The conventional method produces MgAl$_2$O$_4$ layers by the reaction of single-crystal MgO and Al$_2$O$_3$ under high temperature and high stress, yielding layer thicknesses of 10–100 μm. In order to control the growth at a finer scale, pulsed laser deposition (PLD) was employed to produce MgAl$_2$O$_4$ layers in the early growth stage. Below are brief descriptions of the two methods; more details are available in the supplemental material. Previous work have shown that [111]$_{MgAl2O4}$ axis prefers to align with [0001]$_{Al2O3}$ axis during growth, regardless of the initial orientation relationships between the reactants MgO and Al$_2$O$_3$. Therefore (0001) Al$_2$O$_3$ substrates are used for both growth methods, which is convenient for later structure observation.

In the first set of experiments, targeted at early growth stages, amorphous MgO was deposited on polished (0001) single-crystal Al$_2$O$_3$ substrates using PLD. Then the samples were annealed at 900°C for 61 minutes and at 1000°C for 5, 31, 120 and 180 minutes to grow MgAl$_2$O$_4$ layers with 20–400 nm thicknesses, corresponding to sample numbers Cor26, Cor27, Cor29, Cor30 and Cor08 in (Götze *et al.*, 2014). The layer thickness increased at a constant rate, representing the "interface-reaction controlled" early growth stages. The overall morphology for this set of samples is shown in Supplemental Figure 1, showing that at 1000 °C the $Al_2O_3$/$MgAl_2O_4$ interfaces become successively more curved with increasing annealing time. Although the microstructure of the $MgAl_2O_4$ layers show a clear evolution while the annealing time increased, the atomic structure of the interfaces do not show a clear difference in these samples, probably because they are all in the "interface-reaction controlled" early stages. The higher resolution images in Fig. 2 are from sample Cor26, and those in Figs. 3-5 are from sample Cor29.

In the second set of experiments, targeted at late growth stages, $MgAl_2O_4$ layers thicker than 10 μm were produced by the reaction between single-crystals of MgO and $Al_2O_3$ under a uniaxial load of ~30 MPa. The polished (0001) $Al_2O_3$ reactant surface was aligned with (100) of MgO, and annealed in a dry Ar atmosphere. The samples annealed at 1350 °C for 5, 20 and 80 hours were selected for this STEM study, corresponding to sample numbers V27, V26 and CP28 in (Jeřábek *et al.*, 2014). The interface structures do not show clear difference in these samples. Fig. 6 is from sample CP28. Figs. 7 and 8 are from sample V27.

### 2.2. Electron microscopy, image processing, modelling and image simulations

For the 10−100 μm thick $MgAl_2O_4$ layers representing late growth stages, crystal orientation mapping was performed by electron backscatter diffraction (EBSD) analysis using an EDAX Digiview IV EBSD camera mounted on a FEI Quanta 3D field emission gun scanning electron microscope (SEM), at 15 kV accelerating voltage and ~2 nA probe current (Jeřábek *et al.*, 2014). A focused ion beam (FIB) with OmniprobeTM 100.7 micromanipulator, equipped on the same SEM, was used to extract specimens from the selected interface areas. Besides the common FIB lift-out with cross-section geometry, a plan-view lift-out geometry was also applied to extract specimens (Li *et al.*, 2018) in order to study the geometry of "$Al_2O_3$/$MgAl_2O_4$/$MgAl_2O_4$" triple junctions at the $Al_2O_3$/$MgAl_2O_4$ interface from late growth stage samples (Supplemental Figure 2). A 30 kV Ga-ion beam at high current (65 nA - 1 nA) was used for the pre-cut, then 30 kV Ga-ion beam with lower current (1 nA−50 pA) followed by 5 kV and 2 kV with low-current (48 pA−27 pA) were used to thin the specimens to ~ 70nm. A low-kV (0.5−1 kV) argon-milling device was used for final thinning of the specimens to <50 nm thickness. A Nion UltraSTEM $5^{th}$-order aberration-corrected STEM with sub-Å resolution (Krivanek *et al.*, 2011; 2008) was employed to resolve the atomic structure of the interface using accelerating voltage of 100 kV. The probe-forming angle and the inner detector angle for the STEM high angle annular dark field (HAADF) images were approximately 30 and 80 mrad, respectively. To enhance the signal-noise ratio, the experimental images in Fig. 8 were processed as below: STEM image distortions due to instrumental and environmental instabilities were corrected using the IMAGE-WARP procedure ((Rečnik *et al.*, 2005)). Then, background intensity variations visible as horizontal stripes were extracted by filtering low frequency signal using the Gatan DigitalMicrograph software and Wiener filter (Kilaas, 1998). The atomic models for image simulations were built using

CrystalMaker software. STEM HAADF Z-contrast image simulations were performed with Q-STEM software (Koch, 2002), using the experimental electron-optical parameters. More details are available in the supplemental material.

## 3. Results

### 3.1. $Al_2O_3$/$MgAl_2O_4$ interface structure during early growth stages

#### 3.1.1. Nanocrystalline $MgAl_2O_4$ thin layer grown on $Al_2O_3$

$MgAl_2O_4$ thin films with thicknesses from 20 to 400 nm grown between $Al_2O_3$ single crystals and amorphous MgO deposited on the $Al_2O_3$ crystals by PLD document the early growth stages. The $MgAl_2O_4$ thin films show nanocrystalline (nc-) structure, a typical example is shown in Fig. 2a–c, where a ~20 nm thick $MgAl_2O_4$ layer was produced after annealing at 900˚C for 61 minutes. Note the $MgAl_2O_4$ layers grow towards $Al_2O_3$ and MgO simultaneously, however in this research we focus on the interface structure between $MgAl_2O_4$ and $Al_2O_3$.

The STEM-HAADF image intensity is proportional to $\sim Z^{1.7}$, where Z is the atomic number (Z) of each atomic column (Pennycook & Nellist, 2011). Fig. 2c is such a typical Z-contrast image showing the atomic structure of $MgAl_2O_4$ viewed along the $[1\bar{1}0]$ zone axis. In this viewing axis, every second Al column along the $[11\bar{2}]$ and $[112]$ directions has twice the number of atoms as compared to the other Al and Mg columns, and thus shows much brighter intensity in the Z-contrast image. The arrangement of these bright columns (indicated schematically by the blue parallelogram) is a convenient feature for determining the orientations of $MgAl_2O_4$ grains and for assigning interfaces and defects to specific atomic planes in the $MgAl_2O_4$ structure. In Fig. 2b the grain orientations of $MgAl_2O_4$ are again marked out using the blue parallelograms, and $MgAl_2O_4$ grains with diameters of 5 to 20 nm can be observed. The grain boundaries (GBs) between different nano-grains are either coherent Σ3 GBs, referred to as twin boundaries (TBs, marked by white arrow), or low-angle GBs (dashed yellow arrow) and incoherent Σ3 GBs (yellow arrow) that are mostly perpendicular to the reaction interface. The interfaces look uneven and curved. This nanocrystalline configuration is similar in the other PLD-grown $MgAl_2O_4$ thin layers, where the $MgAl_2O_4$ grains grow with two distinct orientations producing contacts including TBs, Σ3 GBs and low-angle GBs. Furthermore, the $MgAl_2O_4$ grain size increases with the increasing layer thickness, as shown in Supplemental Figure 1.

#### 3.1.2. Topotaxial $Al_2O_3$/$MgAl_2O_4$ interfaces: *hcp/ccp* stacking change in the O-sublattice

At the $Al_2O_3$/$MgAl_2O_4$ interface with perfect topotaxial OR, the atomic structures were studied to understand how the lattice transforms from $Al_2O_3$ to $MgAl_2O_4$. The optimum viewing direction for observing the structure of this interface is simultaneously parallel to $[01\bar{1}0]$ in $Al_2O_3$ and $[1\bar{1}0]$ in MgAl$_2$O$_4$. When viewing Al$_2$O$_3$ along the [01$\bar{1}$0] axis, the relative intensities of O and Al columns in the Z-contrast images are similar, because the O-columns are fully occupied, whereas the Al-columns have only a 2/3 occupancy. When viewing MgAl$_2$O$_4$ along the [1$\bar{1}$0] axis, the (111)$_{MgAl2O4}$ planes are occupied by pure Al-layers alternating with Al+Mg mixed layers. For the pure Al layers Al alternately occupies 1/2 or all of the columns, while for the Al+Mg mixed layers Al and Mg occupy 1/2 of their corresponding columns. The fully occupied Al columns in the pure Al-layers give much brighter contrast in the Z-contrast images, as already shown in Fig. 2. This feature was used to identify the interface location. Two different atomic configurations have been observed at dislocation-free interface sections, referred to as "type-A" and "type-B" and shown in the Z-contrast images in Fig. 3 (a1) and (b1). In both Al$_2$O$_3$ and MgAl$_2$O$_4$, cations are located in the center of O polyhedra. Al occupies 2/3 of the O octahedral sites in Al$_2$O$_3$. In MgAl$_2$O$_4$, Al occupies 1/2 of the octahedral sites and Mg occupies 1/8 of the tetrahedral sites. Using all this information, the two interface configurations in Fig. 3 (a1) and (b1) can be simplified in terms of the polyhedral models such as those in Fig. 3 (a2) and (b2), respectively. The two interface configurations have the same topotaxial orientation relationship (OR) with: [1$\bar{1}$0]·(111)$_{MgAl2O4}$ ∥ [01$\bar{1}$0]·(0001)$_{Al2O3}$. The crystallographic OR in reciprocal space for such interface configurations is shown in the corresponding Fast Fourier Transformation (FFT) in Supplemental Figure 5. In both configurations the O-sublattices in Al$_2$O$_3$ follow ABABAB *hcp* stacking, and change to the ABCABC *ccp* stacking in MgAl$_2$O$_4$.

When the oxygen sequence changes from an ABAB *hcp* lattice in Al$_2$O$_3$ to an ABCABC *ccp* lattice in MgAl$_2$O$_4$, there are six possible interface configurations. There are three possible oxygen sequences: A][BA][B(**A**]BC)(ABC), [AB][AB][(A**B**]C)(ABC), and A][BA][BA](CAB)(CAB), where [AB, or BA] and (ABC, or CAB) indicate stacking units in Al$_2$O$_3$ and MgAl$_2$O$_4$, respectively. For each oxygen sequence configuration, the terminating layer of MgAl$_2$O$_4$ at the interfaces can be either the pure Al layer or the Al+Mg mixed layer, therefore there exist a total of six interface configurations. The atomic models and corresponding simulated STEM HAADF images for all six are shown in Supplemental Figure 3. We then compared the two types of experimentally observed interface structure in Fig. 3 (a1) and (b1) with the six possible interface configurations. The matching models are the two in Fig. 3 (a2) and (b2), respectively. The interface structure in experimental image Fig. 3 (a1) is not as sharp as Fig. 3 (b1), which might be due to the 3D configuration of the interface, for instance, the existence of interface steps along the beam direction. However the comparison of atomic structure away from the interface show which is the matching model. The detailed comparison is shown in Supplemental Figure 4.

The stacking sequence from Al$_2$O$_3$ to MgAl$_2$O$_4$ is the same for both cases: BABABABCABCABC. However, the key difference between the two interface configurations is the location of the interface, *i.e.* the position of the shared O plane. The O stacking for type-A interface configuration can be described as [BA][BA][B(**A**]BC)(ABC)(ABC), (**A**) is the O-layer that is shared between the two structures. For type-B interface configuration, the O stacking can be described as B][AB][AB][A(**B**)CA)(BCA)(BC, and (**B**) is the shared O layer. In both structures, the terminating layer of $MgAl_2O_4$ is the pure Al-layer. Although we can not exclude the possible existence of the other four interface configurations, in this paper we only use the observed structures for further modelling.

An alternate way to see the difference between the two interface configurations is by looking at their different bonding configurations. The bonding in $Al_2O_3$ can be considered as a series of zigzag O-Al-O-Al-O chains, forming by interlinked straight O-Al-O sub-chains with alternating directions. For interface structure type-A, the last "O-Al-O" sub-chain in $Al_2O_3$ forms a zigzag O-Al-O-Al-O chain with the last "O-Al-O" sub-chain in $MgAl_2O_4$, as shown by the overlaid red-blue-red-blue-red circles in Fig. 3(a1). The shared O plane (marked by the red arrow) at the interface is located in the middle of this zigzag chain. However, for interface structure type-B, the last "O-Al-O" sub-chain in $Al_2O_3$ near the interface forms a straight O-Al-O-Al-O chain with the last "O-Al-O" sub-chain in $MgAl_2O_4$, as shown by the overlaid red-blue-red-blue-red circles in Fig. 3(b1). The shared O plane is located in the middle of this straight chain. In the Z-contrast images these bonding features are more easily discernible compared to counting O stacking that required separating O columns from the surrounding Al and Mg columns. Therefore, observing whether the bonding follows zigzag or straight O-Al-O-Al-O chains at the interface is a useful way to identify the interface type in lower magnification images such as Fig. 3d, which shows repeated transitions from structure type -A to type-B and again to type-A along the interface. As the $Al_2O_3$/$MgAl_2O_4$ interfaces are located at stacking sequence A for structure type-A and stacking sequence B for structure type-B, the transition from one structure to the other must occur at a surface step with a height equal to an odd number of O stacking sequences. Indeed, the central type-B interface section in Fig. 3d is separated from the type-A interface sections to the left and right of it by steps respectively seven and one $(0001)_{Al2O3}$ O layers high. Fig. 4c shows a closer analysis on the step of one $(0001)_{Al2O3}$ O layer height, which will be further discussed in below section.

### 3.1.3. Partial dislocations at steps of the $Al_2O_3$/$MgAl_2O_4$ interface

At the dislocation-free section of the interface shown in Figs 3 (a-b), the O-sublattice transition is smooth from $Al_2O_3$ to $MgAl_2O_4$. This interface structure cannot be maintained over large areas, because the lattice spacing of the O-sublattice spacing in $Al_2O_3$ is slightly smaller than that in $MgAl_2O_4$ and therefore misfit dislocations are needed to accommodate the lattice mismatch. The lattice mismatch along specific lattice directions is obtained from $(d_{MgAl2O4} - d_{Al2O3})/d_{MgAl2O4}$ (Van Der Merwe, 1978), where $d_{MgAl2O4}$ and $d_{Al2O3}$ are the atomic spacings in $MgAl_2O_4$ and $Al_2O_3$ along the directions examined. In the case at hand, we are interested in the directions $<11\bar{2}>_{MgAl2O4}$ ∥ $<\bar{2}$ 110>$_{Al2O3}$ at the (111)$_{MgAl2O4}$ ∥ (0001)$_{Al2O3}$ interface. d$_{MgAl2O4}$ and d$_{Al2O3}$ are 2.474 Å (Sickafus *et al.*, 1999) and 2.381 Å (Finger & Hazen, 1978), respectively, yielding a mismatch of ~3.76% between the smaller O-sublattice in Al$_2$O$_3$ and the larger one in MgAl$_2$O$_4$.

Misfit dislocations have indeed been observed via STEM, and they are associated with interface steps for both interface types -A and-B, as shown in Figs. 4 (a1) and (b1). Note that each interface step extends over one complete AB stacking sequence in the O-sublattice of Al$_2$O$_3$. Extra O ($\bar{2}$110) planes in Al$_2$O$_3$ (red zigzag lines) are found at such interface steps, terminating at misfit dislocations on the interface. The corresponding polyhedral models of these steps are drawn in Fig. 4 (a2) and (b2), in which Burgers circuits are drawn surrounding the dislocation cores to derive Burgers vectors. Since STEM images are 2D projections of the 3D structure, the component of Burgers vectors parallel to the incident beam direction ([01$\bar{1}$0]$_{Al2O3}$ ∥ [1$\bar{1}$0]$_{MgAl2O4}$) cannot be observed. As the O-O spacings in MgAl$_2$O$_4$ and Al$_2$O$_3$ are very similar, here we use "d" for the O-O spacing along the [11$\bar{2}$]$_{MgAl2O4}$ direction. The observed Burgers vectors for both configurations are 1/3d. Note this is the projection of the full Burgers vector normal to the incident beam direction. For interface type-A in Fig. 4 (a2), the Burgers vector reflects the slip required to move oxygen stacking A to oxygen stacking B. Similarly for interface type-B in Fig. 4 (b2), the Burgers vector reflects the slip required to move oxygen stacking B to oxygen stacking C. In 3D the Burgers vectors from stacking A to B (or B to C) in the *ccp* crystal structure are along the [$\bar{1}$2$\bar{1}$]$_{MgAl2O4}$ or the [2$\bar{1}\bar{1}$]$_{MgAl2O4}$ axes, which are both inclined by 30° with respect to the incident beam direction [1$\bar{1}$0]$_{MgAl2O4}$. With this information we can deduce the full Burgers vectors. If "*a*" is used to describe the edge length of the unit cell of MgAl$_2$O$_4$, the Burgers vectors can be described as (*a*/12)[$\bar{1}$2$\bar{1}$] or (*a*/12)[2$\bar{1}\bar{1}$]. The configuration of these misfit dislocations is very similar to the partial dislocations in cubic structure, e.g. the partial dislocations inside zinc-blende CdTe (Li, Poplawsky *et al.*, 2013; Li, Wu *et al.*, 2013).

The step-associated partial dislocation may appear together with gradual transitions between interface types -A and -B as shown in Fig. 3d. Fig. 4 (c1) is a higher-magnification image of the right side step in Fig. 4d. Here the transition from the interface structure type -B to type-A across a step is actually formed by a partial dislocation associated with a type-B step and accompanied by the gradual insertion of an extra O (0001) plane in Al$_2$O$_3$ (Fig. 4 (c2)). Besides the mismatch of the O-sublattices at the interface plane, there exists an even larger lattice mismatch for O-sublattices along the [0001]$_{Al2O3}$ (or [111]$_{MgAl2O4}$) axis: the O spacing is ~7.145% smaller in Al$_2$O$_3$ than in MgAl$_2$O$_4$. Extra O (0001) planes are needed in Al$_2$O$_3$ to compensate such lattice mismatch during the propagation of MgAl$_2$O$_4$ into Al$_2$O$_3$.

### 3.1.4. Interface between Al$_2$O$_3$ and misaligned MgAl$_2$O$_4$ grains

At the early growth stages, while some MgAl$_2$O$_4$ grains follow the perfect topotaxial OR with Al$_2$O$_3$, some MgAl$_2$O$_4$ grains show a misalignment where the close-packed (0001)$_{Al2O3}$ O basal planes enclose an angle of up to 7° with the close-packed (111)$_{MgAl2O4}$ O planes. A typical example of two MgAl$_2$O$_4$ grains in each situation is shown in Fig. 5a. The left MgAl$_2$O$_4$ grain has its (111) planes exactly parallel to the (0001)$_{Al2O3}$ planes. However, for the right MgAl$_2$O$_4$ grain the (111)$_{MgAl2O4}$ planes have a misalignment of 7° with the (0001)$_{Al2O3}$ planes. The Z-contrast image in Fig. 5b shows the atomic configuration at another inclined interface with 7° misorientation between the (0001)$_{Al2O3}$ and (111)$_{MgAl2O4}$ planes containing "voids" along the interface. The interface sections between the "voids" show relatively ordered bonding configurations, while the sections at the "voids" show a disordered structure. Note the so-called "voids" are not completely empty, but more likely contain vacancies and atoms in irregular bonding. The crystallographic relationships in reciprocal space for the interface configurations in Fig. 5 (a) and (b) are shown in the corresponding FFT in Supplemental Figures 6 and 7, respectively.

Extra ($\bar{2}$110)$_{Al2O3}$ planes in Al$_2$O$_3$ producing misfit dislocations at the interface were identified by counting the close-packed O planes in the *hcp* and *ccp* sublattices of Al$_2$O$_3$ and MgAl$_2$O$_4$. Unlike the partial dislocations formed at the topotaxial interface as Figs 3-4 show, the misfit dislocations in Fig.5 are edge dislocations that have Burgers vectors of full lattice spacing. Fig 5c illustrates that for such inclined interfaces, close-packed oxygen lattices do not transform smoothly as those at perfect topotaxial interfaces, instead it requires a rearrangement of O atoms from *hcp* to *ccp* sublattices.

Such misaligned interfaces are only observed during the early growth stages, the possible formation mechanism and their impact on interface migration will be discussed in section 4.2.

### 3.2. Al$_2$O$_3$/MgAl$_2$O$_4$ Interface structure during late growth stages

#### 3.2.1. Crystallographic OR and microstructure evolution during layer growth

MgAl$_2$O$_4$ layers thicker than 10 μm were grown between single crystal Al$_2$O$_3$ and MgO precursors at 1350°C and under a uniaxial load of ~30 MPa. A typical microstructure of the thick MgAl$_2$O$_4$ layers is shown in the EBSD inverse pole figure (IPF) map in Fig. 6. The MgAl$_2$O$_4$ product layer is comprised of two sub-layers. The sub-layer on the side of MgO makes up about 25% of the total layer thickness and is comprised of columnar grains with their long axis perpendicular to the MgO/MgAl$_2$O$_4$ interface. In this sub-layer the MgAl$_2$O$_4$ grains have a simple topotaxial OR with MgO with {001}$_{MgAl2O4}$ ∥ {001}$_{MgO}$. The columnar grains in the MgO-orientated sub-layer show small misorientations of about 2° between adjacent grains (Li, Griffiths *et al.*, 2016). The sub-layer on the side of Al$_2$O$_3$ makes up about 75% of the layer thickness. In this sublayer the MgAl$_2$O$_4$ grains show a topotaxial OR with Al$_2$O$_3$ that is [1$\bar{1}$0]·(111)$_{MgAl2O4}$ ∥ [01$\bar{1}$0]·(0001)$_{Al2O3}$. Within this OR there exist two twin variants (dark and light green). The boundaries between these two twin variants are Σ3 GBs, including incoherent ones and coherent ones, the latter are TBs.

In these thick layers, the $MgAl_2O_4$ grains of the $Al_2O_3$-orientated sub-layer show a successive grain-size increase from the original $Al_2O_3$/MgO interface towards the final $Al_2O_3$/$MgAl_2O_4$ interface (Fig. 6). The Σ3 boundaries in $MgAl_2O_4$ are oriented approximately perpendicular to the original $Al_2O_3$/MgO interface during the early growth stages, and most of them successively change orientation towards parallel to the original $Al_2O_3$/MgO interface during later growth stages. Some of the small grains near the original $Al_2O_3$/MgO interface deviate by up to 7° from the typical OR. This misalignment feature is similar to what is observed during the initial growth stage (Fig. 3). The larger grains that are closer to the $Al_2O_3$/$MgAl_2O_4$ interface show much better alignment with the topotaxial OR (Jeřábek et al., 2014).

### 3.2.2. Continuous steps at $Al_2O_3$/$MgAl_2O_4$ Interface intersecting with Σ3 GBs

The Σ3 GBs between $MgAl_2O_4$ grains have irregular shapes on the scale of EBSD maps (Fig. 6). At higher magnifications, however, the Σ3 GBs are clearly formed by facets with specific orientations (Fig. 7). The $Al_2O_3$/$MgAl_2O_4$ interface plane typically changes its orientation at triple junctions, where it intersects with $MgAl_2O_4$ Σ3 GBs. Several examples of such triple junctions are shown in Supplemental Figure 2.

Fig. 7 shows the detailed structure of such an $MgAl_2O_4$-$MgAl_2O_4$-$Al_2O_3$ triple junction, where the $Al_2O_3$/$MgAl_2O_4$ interface changes its orientation abruptly (Fig. 7a). The two $MgAl_2O_4$ grains in Fig. 7 have a twin OR. The interfaces on the two sides of the triple junction are comprised of interface facets at $\{111\}_{MgAl2O4}\|\{0001\}_{Al2O3}$ planes and steps that run in opposite directions (Fig. 7b-c). The boundary between the two $MgAl_2O_4$ grains is formed by alternating segments of incoherent Σ3 GBs and TBs (Fig. 7d-e). The TBs are always parallel to the $(111)_{MgAl2O4}$ planes, also called coherent Σ3$\{111\}$ GBs. TBs are found parallel to the $Al_2O_3$/$MgAl_2O_4$ interface facets at the segments between steps. In most of the areas, the cation layers in the two $MgAl_2O_4$ grains align with each other across the Σ3 incoherent GBs. The crystallographic relationship in reciprocal space for Fig. 7 is shown in the corresponding FFT in Supplemental Figure 8.

For both interface structures shown in Fig.7b and c, there are small misalignments between the $(111)_{MgAl2O4}$ planes and the $(0001)_{Al2O3}$ planes: 1.2° for Fig. 7b, 0.8° for Fig. 7c. Due to such small misalignments, the interface has to crosscut the stacking sequence in $Al_2O_3$, resulting in an alternation between structure types -A and -B at the interface. One such example is shown on the left side of the triple-junction in Fig. 7(d2): the interface structure changes from type-A to type-B towards the right direction. Note that the atomic structure at the interface planes is not always sharp, which might be due to the 3D structure of the interface or surface contamination. However, by comparing the columns 2-3 atomic layers away from the interface, we are able to distinguish the interface types (Fig. 7(d2)).

At the MgAl$_2$O$_4$-MgAl$_2$O$_4$-Al$_2$O$_3$ triple-junction where the two interfaces meet, the interface structure at the left side follows type-B while the right side follows type-A. Right at the triple junction, the cation layers in the two MgAl$_2$O$_4$ grains are well aligned, with one O plane height difference between the locations of the Al$_2$O$_3$/MgAl$_2$O$_4$ interfaces at the left and right sides. These observations are important for modeling the formation of MgAl$_2$O$_4$ in Section 4.

### 3.2.3. Atomic structure of twin boundaries in MgAl$_2$O$_4$

TBs in MgAl$_2$O$_4$ are observed in relatively long flat sections, parallel to the (111)$_{MgAl2O4}$ planes. Two such segments are shown in Fig. 7e. When TBs terminate at Σ3 GBs, there exists strain. However the middle portions of the TBs show identical atomic structure, suggesting that they are generated in a similar way. The atomic configurations of a TB are shown in Fig. 8(a1) and (b2), viewed along the $[1\bar{1}0]$ and the $[11\bar{2}]$ zone axes, respectively. The twin structure is mirror-symmetric across the (111)$_{MgAl2O4}$ plane. In Z-contrast STEM images, the twin planes show a lower intensity than the surrounding planes, which might be caused by lower atomic density compared to the normal MgAl$_2$O$_4$ structure or vacancy-induced strain (Findlay *et al.*, 2011). In order to determine the atomic structure of the TBs, experimental Z-contrast images in Fig. 8(a1) and (b1) have been processed to enhance the contrast features, resulting in Fig. 8(a2) and (b2) respectively. Based on the surrounding atomic structure, the TB structure was reconstructed. Exactly at the TB the columns have twice the O-O spacing of the normal (111)$_{MgAl2O4}$ plane. The two MgAl$_2$O$_4$ grains symmetrically meet at the TB with completed (111) pure Al layers, followed by partially occupied Al+Mg mixed layers, where only the Mg-tetrahedral sites are occupied. To identify the nature of the columns along the TB plane, several different arrangements were considered, including Al-, Mg-, and O-atoms on the twin plane. As the atoms directly adjacent to the twin planes are Mg atoms and Mg-O bonding is much more energetically stable than Mg-Al or Mg-Mg bonding, the most likely atoms on the TB plane are O atoms. Correspondingly TB models Fig. 8 (a4) are built and STEM Z-contrast image simulations Fig. 8 (a3) were performed, showing a good match with the experimental image in Fig. 8 (a2). The consistency of this TB model was verified in a perpendicular projection, shown in Fig. 8 (b1-4). Note the absolute intensity on the twin planes in experimental images is not as uniform as in the simulated images, which might be due to variation of O occupancy or local strain.

### 4. Discussion

### 4.1. Partial dislocation glide and interface migration

The O-sublattices of Al$_2$O$_3$ and MgAl$_2$O$_4$ may be described as *hcp* and *ccp* structures, respectively. The classical geometry of an interface between a *hcp* and a *ccp* lattice has been described by Porter et al. (Porter *et al.*, 2009) with simple sketches. A glissile interface is formed by an array of Shockley partial dislocations, which are produced during the rearrangement of the stacking sequence from ABAB in the *hcp* lattice to ABCABC in the *ccp* lattice. Our results show that this classical geometry not only applies to the case of a simple crystal lattice containing only one element, such as in metals, but also provides a good description for oxides, with the O-sublattice following the *hcp* to *ccp* transformation, while the cations occupy different sets of interstitial sites of the O-sublattices.

It is worth mentioning that Displacement Shift Complete (DSC) lattice model has been adapted for illustrating steps on a phase boundary between a hexagonal and a cubic lattice by Professor H. Föll (Chapter 8.3.1, Föll, online book). With the sketch of steps, he pointed out "steps can be incorporated without problems and without dislocations as long as the step height comes in multiples of 3 (in units of the translation vectors of the coincident site lattice (CSL))". However when the step height comes in multiples of 2 or 1 in units of the translation vectors of the CSL, dislocations have to be involved. This is consistent with our observation.

We find that oxygen layers at both stacking positions A and B can form the termination plane of $Al_2O_3$ at the $Al_2O_3$/$MgAl_2O_4$ interface, leading to two interface structures, type-A and type-B. Apart from the bonding configuration at the interface planes, these two structural types are very similar, and both show partial dislocations that are associated with steps in the $Al_2O_3$/$MgAl_2O_4$ interface. These partial dislocations at the $Al_2O_3$/$MgAl_2O_4$ interface have not been resolved in previous studies on similar systems (Carter & Schmalzried, 1985; Sieber *et al.*, 1996; Hesse *et al.*, 1994). This is probably due to the limited resolution of earlier transmission electron microscopes without aberration-correction. Moreover, the Burgers vector of partial dislocations is smaller than the Burgers vector of a full dislocation. As a consequence, the strain associated with a partial dislocation is comparatively small and makes structure determination difficult.

The out-of-plane mismatch of the O-sublattices along the directions $[0001]_{Al2O3}$ is even larger than the in-plane mismatch of the O-sublattices at the interface plane. This larger out-of-plane mismatch can be compensated by extra $(0001)_{Al2O3}$ planes. This is likely the reason for a small misalignment of ~1° which is commonly observed between the $(111)_{MgAl2O4}$ and $(0001)_{Al2O3}$ planes, leading to an alternation of interface structure types -A and -B, as observed in both early (Fig. 3d, Fig. 4c) as well as late (Fig. 7b-c) growth stages.

From the observed atomic structure of the $Al_2O_3$/$MgAl_2O_4$ interface, we infer that the interface propagates by the glide of misfit dislocations and the migration of the associated interface steps along the interface. The misfit partial dislocations are located at the terminating $(0001)_{Al2O3}$ planes at the $Al_2O_3$/$MgAl_2O_4$ interface, and they are associated with steps at the interface, where one step is two closed-packed O-layers high. The glide of the partial dislocations can be accomplished by two elementary moves as illustrated in Fig. 9, in which interface structure type-B is used to illustrate the gliding process. The gliding process for interface structure type-A is similar.

Two consecutive moves of glide transform structure 1 to structure 2 and then to structure 3. Structure 3 is in fact identical to structure 1 except that the partial dislocation and the associated interface step have moved sideways by a distance of two $AlO_6$ octahedrons (see Fig. 9a(1-3)). The glide process is accomplished in the following way. During the first elementary move of glide, O column $O_1$ detaches from the *hcp* O-sublattice of $Al_2O_3$, glides to the left, and joins the *ccp* O-sublattice of $MgAl_2O_4$ (from Fig. 9(a1) to (a2)). Meanwhile, the rearrangement and exchange of cations takes place. The two $Al^{3+}$ columns in $Al_2O_3$ on the upper and lower right of the $O_1$ column in Fig. 9(a1)) are replaced by the dislocation core, and the two $Al^{3+}$ columns in $MgAl_2O_4$ on the upper- and lower left of the $O_1$ column are formed at the previous location of the dislocation core (Fig. 9(a2)). Following this process, a structural unit of $Al_2O_3$ in $Layer_1$ containing four $Al^{3+}$ cations is transformed into a structural of $MgAl_2O_4$ containing six $Al^{3+}$ cations. Simultaneously, a structural unit of $Al_2O_3$ in $Layer_2$ containing four $Al^{3+}$ cations is transformed into a structural unit of $MgAl_2O_4$ containing three $Al^{3+}$ cations. This is followed by a second glide move, in which the O column $O_2$ detaches from the *hcp* O-sublattice of $Al_2O_3$, glides to the left, and joins to the *ccp* O-sublattice of $MgAl_2O_4$. The two Al columns on the upper and lower right of the O column $O_2$ pertaining to $Layer_1$ and $Layer_2$ of $Al_2O_3$ are replaced by the dislocation core (Fig. 9(a2)), and one Al column and two Mg columns pertaining to $Layer_1$ and $Layer_2$ of $MgAl_2O_4$ are formed at the previous location of the dislocation core (Fig. 9(a3)). With this operation, a structural unit of $Al_2O_3$ in $Layer_1$ containing four $Al^{3+}$ cations is transformed into a structural unit of $MgAl_2O_4$ containing three $Al^{3+}$ cations, and a structural unit of $Al_2O_3$ in $Layer_2$ containing four $Al^{3+}$ cations is transformed into a structural unit of $MgAl_2O_4$ containing six $Mg^{2+}$ cations. Such partial dislocation migration caused by glide of atomic columns in the dislocation cores has also been observed in CdTe with *ccp* structure by in-situ STEM (Li, Zhang *et al.*, 2016; Li *et al.*, 2017).

After the two elementary glide moves in $Layer_1$, a structural unit of $Al_2O_3$ containing eight $Al^{3+}$ cations is transformed into a structural unit of $MgAl_2O_4$ containing nine $Al^{3+}$ cations, and in $Layer_2$, a structural unit of $Al_2O_3$ containing eight $Al^{3+}$ cations is transformed into a structural unit of $MgAl_2O_4$ containing three $Al^{3+}$ and six $Mg^{2+}$ cations. The net mass transfer is thus: 16 $Al^{3+}$ → 12 $Al^{3+}$ + 6 $Mg^{2+}$. Considering long-range transfer of $Mg^{2+}$ and $Al^{3+}$ cations across the $MgAl_2O_4$ layer, the net mass transfer can also be described as: 16 $Al^{3+}$ + 6 $Mg^{2+}$ ⇌ 12 $Al^{3+}$ + 6 $Mg^{2+}$ + 4 $Al^{3+}$. The sixteen $Al^{3+}$ are derived from $Al_2O_3$ and six $Mg^{2+}$ are delivered from the $MgAl_2O_4$/MgO interface via long-range diffusion. After a structure unit of $Al_2O_3$ containing sixteen $Al^{3+}$ has transformed into a structural unit of $MgAl_2O_4$ containing twelve $Al^{3+}$ and six $Mg^{2+}$, the remaining 4 $Al^{3+}$ cations are liberated and migrate to the $MgAl_2O_4$/MgO interface via long-range diffusion, where they take part in the transformation of MgO to $MgAl_2O_4$.

The change in the stacking sequence of the O-sublattice during the transformation of $Al_2O_3$ to $MgAl_2O_4$ may be envisioned like the motion of a zipper along the interface, which is accompanied by the rearrangement of $Mg^{2+}$ and $Al^{3+}$ cations. This mechanism repeats along larger segments of the $Al_2O_3$/$MgAl_2O_4$ interface and facilitates its propagation into the reactant $Al_2O_3$. Note the gliding mechanism underlying the propagation of the $Al_2O_3$/$MgAl_2O_4$ interface is fundamentally different from the migration mechanism of the MgO/$MgAl_2O_4$ interface on the other side of the growing $MgAl_2O_4$ layer, where the *ccp*/*ccp* MgO/$MgAl_2O_4$ interface migrates by the climb of misfit dislocations in $MgAl_2O_4$ (Li, Griffiths *et al.*, 2016).

The exchange of cations at the $Al_2O_3$/$MgAl_2O_4$ interface is necessarily linked to the transformation of MgO to $MgAl_2O_4$ at the MgO/$MgAl_2O_4$ interface through long-range diffusion. It has been shown by (Abart *et al.*, 2016) that the overall kinetics of $MgAl_2O_4$-layer growth at the contacts between MgO and $Al_2O_3$ is controlled by the coupling of long-range diffusion of $Al^{3+}$ and $Mg^{2+}$ and interface reaction, where the latter comprises all processes that are localized at the MgO/$MgAl_2O_4$ and the $Al_2O_3$/$MgAl_2O_4$ interfaces. Based on systematic deviations from equilibrium Al/Mg partitioning between the phases at the MgO/$MgAl_2O_4$ and the $Al_2O_3$/$MgAl_2O_4$ interfaces, (Abart *et al.*, 2016) inferred relatively low mobility for the MgO/$MgAl_2O_4$ interface and relatively high mobility for the $Al_2O_3$/$MgAl_2O_4$. This is in line with migration of the MgO/$MgAl_2O_4$ interface by dislocation climb, which is non-conservative with respect to the O sub-lattice and requires delivery of oxygen atoms through the formation of Schottky defects in the reactant MgO and transport of the oxygen atoms to the edge dislocations by volume diffusion, both of which are dissipative processes. In contrast, the glide mechanism underlying the migration of the $Al_2O_3$/$MgAl_2O_4$ interface is conservative with respect to the O sub-lattice and less sluggish.

**4.2. Comparison of early and late growth stages**

The partial dislocations and the associated steps in the $Al_2O_3$/$MgAl_2O_4$ interface are observed during both the early and the late growth stages, suggesting that growth by dislocation glide occurs at all growth stages. However, there are some fundamental differences between the early and late growth stages. Firstly, the $MgAl_2O_4$ grains from the early stages are more columnar, and hence the GBs are oriented more perpendicular to the $Al_2O_3$/$MgAl_2O_4$ interface compared to the GBs from the late growth stages. Secondly, at the early stages, the crystallographic orientations of some $MgAl_2O_4$ grains deviate by up to 7° from perfect topotaxy with $Al_2O_3$ so that the close-packed $(111)_{MgAl2O4}$ and $(0001)_{Al2O3}$ planes are inclined to one another. In the conventionally grown $MgAl_2O_4$ layers, which are more than 10 μm thick, such large deviations from the perfect topotaxy relationship were only observed within the nanocrystalline grains that were formed near the original $Al_2O_3$/MgO interface during the early growth stages. Within the sub-layer that grew from the original $Al_2O_3$/MgO contact towards $Al_2O_3$, a continuous increase in the size of the $MgAl_2O_4$ grains is observed, as shown in Fig. 6. The larger $MgAl_2O_4$ grains close to the $Al_2O_3$/$MgAl_2O_4$ interface show ORs close to the ideal *hcp-ccp* topotaxial relationship to $Al_2O_3$ (Jeřábek *et al.*, 2014). Hence during layer growth, among the many small MgAl$_2$O$_4$ grains that initially formed at the original Al$_2$O$_3$/MgO contact, a subset of them with well-aligned lattice OR to Al$_2$O$_3$ preferably continue to evolve, whereas the others cease to grow.

A 5-7° misalignment between the respective close-packed planes in MgAl$_2$O$_4$ and Al$_2$O$_3$ at the early growth stage has been reported previously, and possible origins have been discussed. It has been proposed that this inclined interface structure can be explained because it allows four $(1\bar{1}\bar{4})$ Al$_2$O$_3$ planes to match the spacing of five (400) planes in MgAl$_2$O$_4$ (Sieber *et al.*, 1996; Hesse *et al.*, 1994). Although this geometric correspondence of lattice plane spacings exists, the model does not explain how the atoms can attain a low energy structure at the interface. Fig.5 (b) demonstrates that although some planes in Al$_2$O$_3$ coincide with some planes in MgAl$_2$O$_4$, the atomic structure in-between these coincident planes is messy, as shown by the periodic "voids" at the interface. Meanwhile edge dislocations are present for such interfaces. In order for such an interface to migrate, all the atoms need to be rearranged, and edge dislocations need to climb, which requires more energy than dislocation gliding. Another explanation inferred that the inclined interfacial structure forms in order to accommodate in-plane lattice mismatch at the interface (Carter & Schmalzried, 1985). The inclined interfacial structure indeed helps to accommodate lattice mismatch at the interface, but even the observed maximum angle (7°) would not be large enough to compensate for the entire 3.76% mismatch between Al$_2$O$_3$ and MgAl$_2$O$_4$ lattices. Therefore misfit dislocations are still needed, which explains why we observe edge dislocations at inclined interfaces in Fig. 5.

Our results show that this misalignment only exists during the early growth stages, and is absent in the late growth stages, suggesting such structural change results from the change in the growth mechanism as the growth proceeds. Both edge and partial dislocations are observed in the early growth stage, while mainly partial dislocations are observed at the late growth stage. Because edge dislocations climb and partial dislocations glide, such evolution of interface structure from early to late growth stages indicates the evolution of the migration mechanism from mixed climbing+gliding towards mostly gliding. Note the glide and climb mechanisms refer to the O-sublattices, and Al-Mg cation interdiffusion is necessary for both.

As the thickness of the growing MgAl$_2$O$_4$ layer increases, the growth switches from interface-reaction-controlled to diffusion-controlled (Abart *et al.*, 2016). MgAl$_2$O$_4$ layers are grown towards the Al$_2$O$_3$ and MgO precursors simultaneously. During early growth stages when the growing MgAl$_2$O$_4$ layer is very thin, diffusion distances between the MgAl$_2$O$_4$/Al$_2$O$_3$ and the MgAl$_2$O$_4$/MgO interfaces are very short. As a consequence, diffusion is very efficient, Al$^{3+}$ and Mg$^{2+}$ cations are abundant at interfaces, and interface reaction is the limiting factor for the overall growth rate. Chemical potential jumps exist at the MgAl$_2$O$_4$/Al$_2$O$_3$ and the MgAl$_2$O$_4$/MgO interfaces for both the MgO and the Al$_2$O$_3$ components, providing a local thermodynamic driving force for interface motion. In this situation, the MgAl$_2$O$_4$/Al$_2$O$_3$ interface forms a configuration that maximizes its mobility. A high interface mobility is ensured by a high density of interface dislocations. The local thermodynamic driving force outweighs the energetically less favorable interface configuration and the energetically expensive dislocation climb associated with the migration of the inclined interface structure.

However during the late growth stages, when the $MgAl_2O_4$ layer becomes thicker, diffusion becomes less efficient and eventually becomes rate limiting. During diffusion-controlled growth, the local thermodynamic driving force at the $MgAl_2O_4/Al_2O_3$ interface diminishes. As a consequence, the interface velocity and mobility decrease, and the interface tends to organize itself into a low-energy configuration with the O-sublattices of $MgAl_2O_4$ and $Al_2O_3$ well aligned and lattice mismatch accommodated by partial dislocations. At the late growth stage the $MgAl_2O_4/Al_2O_3$ interface migrates by the glide of partial dislocations in O-sublattices. The continuous interface steps at the late stage (Fig. 7) are strong proof of the gliding mechanism. This is fundamentally different from the situation at the $MgO/MgAl_2O_4$ ccp/ccp interface for which gliding growth is preferred at the early stage while climbing growth is dominant at the late growth stage (Sieber, Hesse, Werner *et al.*, 1997; Li, Griffiths *et al.*, 2016).

Under the slower growth regime $MgAl_2O_4$ grains compete, with those that produce the lowest-energy configurations winning out at the $MgAl_2O_4/Al_2O_3$ interface. Energy minimization between neighboring $MgAl_2O_4$ grains produces low-energy $\Sigma 3$ boundaries, as shown in Fig. 7. The mechanism of $MgAl_2O_4$ growth under this regime is illustrated in Fig. 10 according to the observed structure in Fig. 7. At the $Al_2O_3/MgAl_2O_4/MgAl_2O_4$ triple junction the $Al_2O_3/MgAl_2O_4$ interface structures at the left and right side have interface structures -B and -A, respectively. The $MgAl_2O_4$ grain on the left side migrates towards the right, while the grain on the right side migrates towards the left, both by dislocation glide. Where the two $MgAl_2O_4$ grains meet, one grain grows faster into $Al_2O_3$ than the other. In Fig. 7, the grain on the right side grew faster than the one on the left, but this selection of faster growth does not seem to be due to the interface type, as the interfaces on either side of the triple point consist of segments of both structures -A and -B. There is no clear correlation with an individual twin orientation of the $MgAl_2O_4$ either, as there is no clear selection of either orientation for the grains from the EBSD maps. The possible reason for the selection of growth direction might be due to external forces such as stress, because in the EBSD map (Fig. 6) most of the GBs show a preferred inclined angle towards the upper left direction at the microscale.

Nearby $MgAl_2O_4$ grains of opposite twin orientations produce $\Sigma 3$ GBs. Across the incoherent $\Sigma 3$ GBs the Mg+Al mixed layers are mostly well aligned from one grain to the other. The translation state between the two nearby $MgAl_2O_4$ grains at (111) planes produce TBs that run parallel with the $MgAl_2O_4/Al_2O_3$ interface. Note that there exists strain when a TB merges into a GB, as there is one more O (111) plane at the TB, this is not shown in the simple sketch in Fig. 10. The main feature of these TBs is their mirror symmetry, which is fundamentally different from the atomic structure of chemically induced TBs formed in natural spinel, such as reported from Be-doped $MgAl_2O_4$ (Daneu *et al.*, 2007; Drev *et al.*, 2013), where the twinning is accomplished by 180° rotation around the [111] axis. It is also different from the twin structure reported in other spinel structures, such as twins in magnetite (Gilks *et al.*, 2016). Furthermore, some reported natural (111) twin boundaries in spinel are chemically induced by incorporation of $Be^{2+}$ that replaces $Mg^{2+}$ on the tetrahedral sites in the local *hcp* stacking, as demonstrated by (Daneu *et al.*, 2007; Drev *et al.*, 2013). Here, the TB is free from any dopant. Similar TB structures would be expected in other pure spinels (Hornstra, 1960).

**4.3. Impact for further work**

The above results illustrate how understanding interface structures can lead to an understanding of solid-state reaction kinetics. The present atomic interface structures also supply a good basis for further theoretical work such as density functional theory (DFT) and molecular dynamic (MD) calculations. Furthermore, this new understanding of the interface structure and migration mechanism in the present materials sheds light on many other systems that involve *hcp*/*ccp* interface transformations, not only oxide/oxide interfaces as discussed in this paper and other oxide interfaces (Zhou *et al.*, 2017), but also metal/oxide interfaces such as the $Al_2O_3$/Al interface (Pilania *et al.*, 2014), or metal/metal interfaces such as in Ru/Pt core/shell nanoparticles (Hsieh *et al.*, 2013). The *hcp*/*ccp* transformation mechanism for a solid-state reaction revealed in this paper might also apply to phase change of oxides or metals under extreme conditions (temperature, pressure, strain, etc). For instance, iron oxide can transfer between $Fe_3O_4$ (magnetite, *ccp* O-sublattice in inverse spinel structure), γ-$Fe_2O_3$ (maghemite, *ccp* O-sublattice in metastable cubic phase) phase and α-$Fe_2O_3$ (hematite, *hcp* O-sublattice similar to α-$Al_2O_3$). The transitions from the first two structures to the α-$Fe_2O_3$ structure both involve migration of a *hcp*/*ccp* interface (Kachi *et al.*, 1963; Genuzio *et al.*, 2016), which happen under high temperature annealing. Another example is the metal titanium, which changes phase from *hcp* to *ccp* during room temperature rolling of thin titanium plates (Wu *et al.*, 2016). It has also been reported that similar *hcp*/*ccp* phase changes can also happen in ambient conditions for gold square sheets via surface ligand exchange or surface coating (Fan *et al.*, 2015). Another common type of *hcp*/*ccp* phase change is the wurtzite/zinc-blende phase change in semiconductors such as Si, InP, CdTe, $Cd_xZn_{1-x}S$, etc (Liu *et al.*, 2013; Li *et al.*, 2014; Wood & Sansoz, 2012; Bakke *et al.*, 2011; Akopian *et al.*, 2010; Bao *et al.*, 2008; Murayama & Nakayama, 1994). Therefore this new fundamental understanding of the $Al_2O_3$/$MgAl_2O_4$ interface migration mechanism might greatly promote understanding of growth and phase changes in many other materials.

**5. Conclusions**

$MgAl_2O_4$ layers have been grown by reaction between MgO and $Al_2O_3$, and the atomic structure of the $Al_2O_3$/$MgAl_2O_4$ interfaces investigated in order to interpret the interface migration mechanism during different growth stages. During the early growth stages some $MgAl_2O_4$ grains follow the topaxial OR with Al$_2$O$_3$ [1$\bar{1}$0]·(111)$_{MgAl2O4}$ ∥ [01$\bar{1}$0]·(0001)$_{Al2O3}$, while others show up to 7° deviations from this topotaxial OR. During later stages of MgAl$_2$O$_4$ layer growth, both the ORs and the atomic structure at the Al$_2$O$_3$/MgAl$_2$O$_4$ interfaces become more organized. At interfaces with a perfect topotaxial OR, the stacking of the O-sublattice changes from *hcp* in Al$_2$O$_3$ to *ccp* in MgAl$_2$O$_4$, with one O layer shared by the *hcp* and *ccp* O-sublattices. Depending on whether the position of the shared O layer is at stacking position A or B in Al$_2$O$_3$, there exist two interface structures, type -A and -B, with exactly the same OR but different atomic bonding at the interfacial plane. The in-plane lattice mismatch at the Al$_2$O$_3$/MgAl$_2$O$_4$ interfaces is accommodated by partial dislocations, which are associated with interface steps with a height of two O stacking layers in Al$_2$O$_3$. Based on the comparison of the interface structure at early and late growth stages, we inferred that both climbing and gliding growth occurs at the early stage, however the gliding growth mechanism is almost exclusively favoured at the late growth stage. The gliding growth occurs by glide of the partial dislocations in O-sublattices along the Al$_2$O$_3$/MgAl$_2$O$_4$ interface plus the accompanying inter-diffusion of Al$^{3+}$ and Mg$^{2+}$ cations. The repetition of this gliding process involving many layers can be envisioned as a zipper-like layer-by-layer transformation. When two neighbouring MgAl$_2$O$_4$ grains with twinned orientations coalesce during growth, alternating TB and incoherent Σ3 GB segments are formed. The switch of growth mechanism at Al$_2$O$_3$/MgAl$_2$O$_4$ *hcp*/*ccp* interfaces from climbing in the early stage to gliding in the late stage is opposite to the situation at the MgO/MgAl$_2$O$_4$ *ccp*/*ccp* interface, for which growth by glide is preferred at the early stage while growth by climb is dominant at the late stage, showing the significant influence of interface structure on the growth mechanism. This fundamental understanding of the migration of Al$_2$O$_3$/MgAl$_2$O$_4$ oxides interface offers significant benefits for the understanding of many other interfaces with similar *hcp*/*ccp* lattice configurations.

**Acknowledgements**  This research was funded by the European Union's Horizon 2020 research and innovation programme under the Marie Skłodowska-Curie grants No. 656378 – Interfacial Reactions (CL) & No. 655760 – Digiphase (TJP), and the Austrian Science Fund (FWF): I1704-N19 in the framework of the research group FOR741-DACH (GH). CL and TJP acknowledge the support from Max Planck Institute. CL acknowledges Dr. Wilfried Sigle and the two anonymous reviewers for their helpful comments. The FEI Quanta 3D FEG-SEM instrument is supported by the Faculty of Geosciences, Geography and Astronomy, and the Nion UltraSTEM is supported by the Faculty of Physics, University of Vienna.

**Figures**

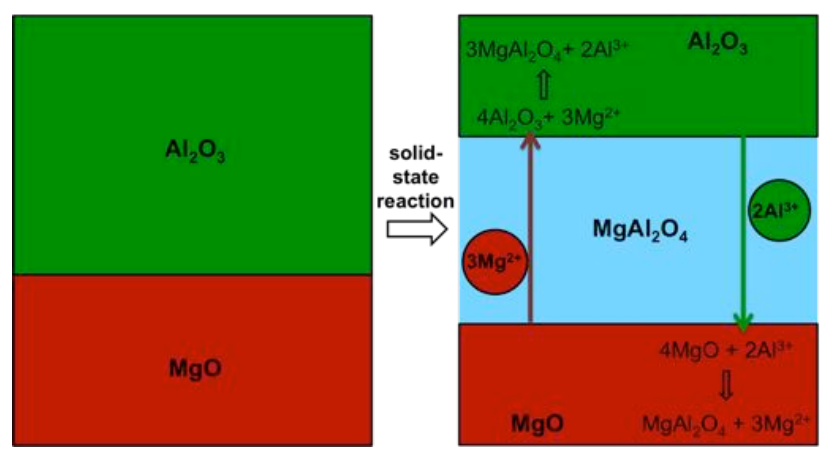

**Fig. 1.** A sketch showing the exchange of cations between MgO and $Al_2O_3$ forming a $MgAl_2O_4$ interlayer during solid-state reaction.

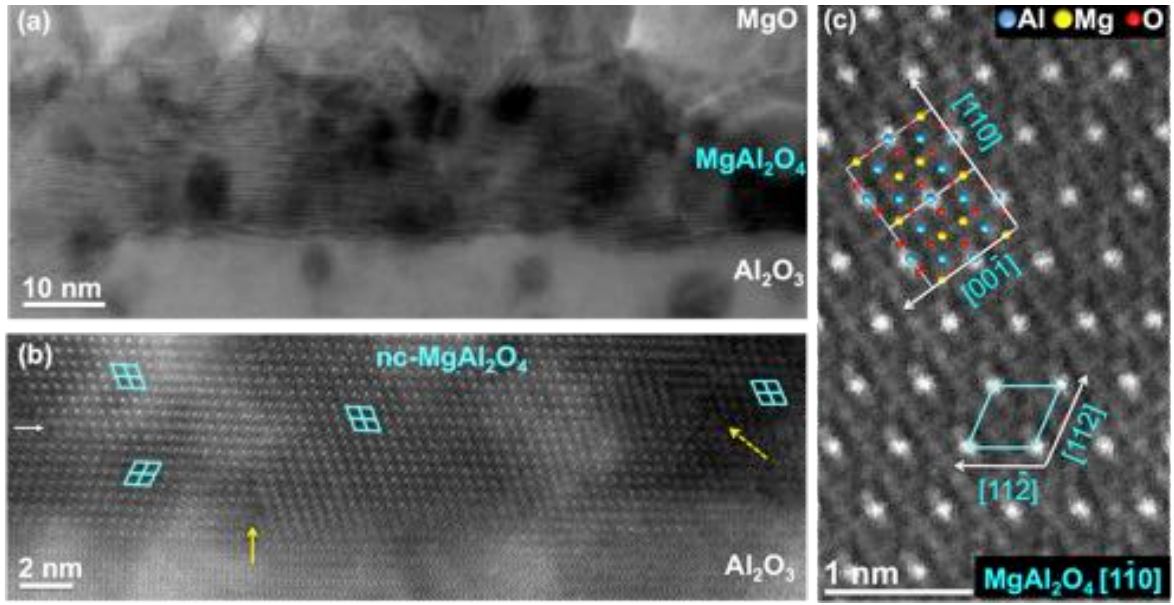

**Fig. 2.** Morphology of nanocrystalline (nc-) $MgAl_2O_4$ layers. (a) STEM-bright field (BF) image and (b) higher-magnification HAADF Z-contrast images show $MgAl_2O_4$ layer grown at 900°C for 61 minutes. (c) STEM Z-contrast image shows atomic structure of $MgAl_2O_4$. The corners of the blue parallelogram mark the positions of four "bright" Al columns that have double intensity along the $[1\bar{1}0]$ axis, which is a convenient feature that highlights the orientation of the $MgAl_2O_4$ lattice. The nanocrystalline (nc-) $MgAl_2O_4$ layer is full of TB (white arrow), Σ3 GB (yellow arrow) and small angle GB (dashed yellow arrow).

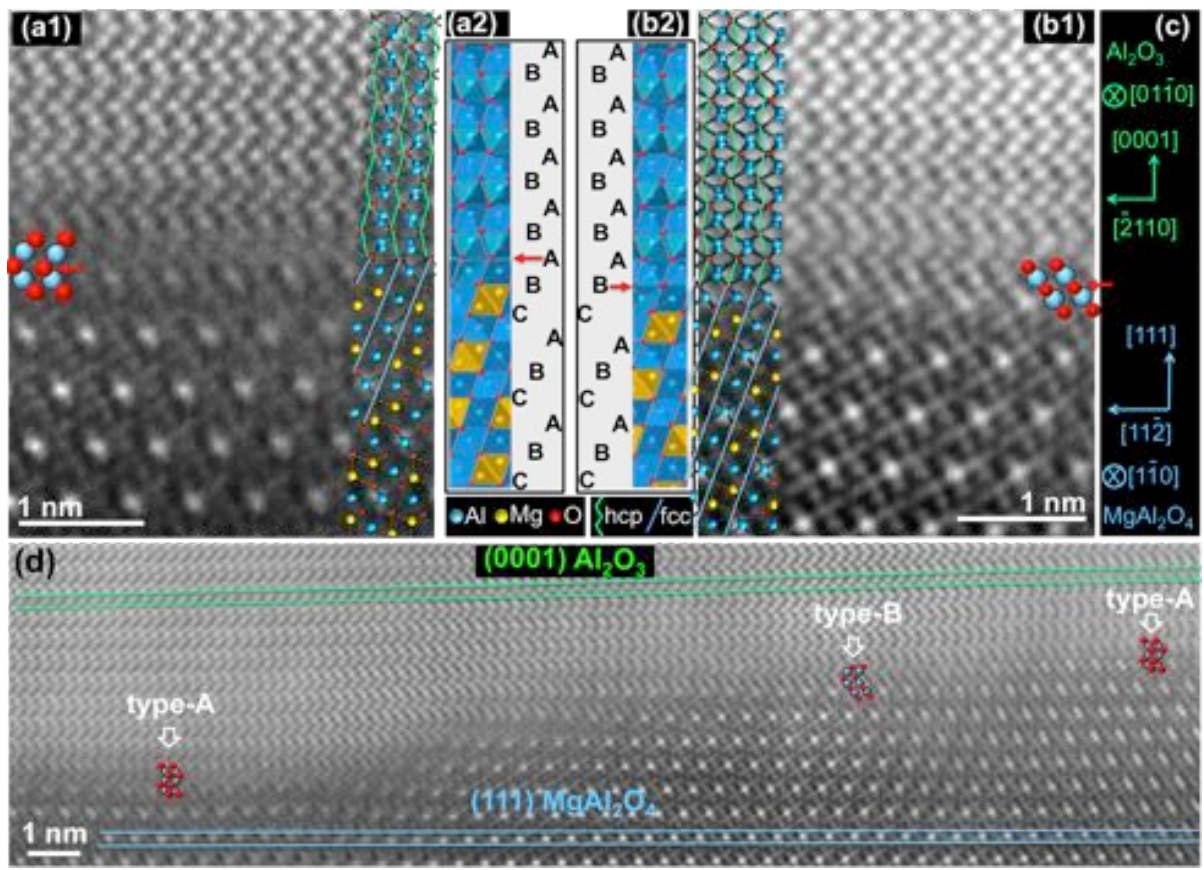

**Fig. 3.** STEM Z-contrast images showing two different atomic configurations of $Al_2O_3$/$MgAl_2O_4$ interfaces at dislocation-free sections: (a1) structure type-A and (b1) structure type-B with the corresponding models in (a2) and (b2), with Al (blue octahedra) and Mg (yellow tetrahedra) coordination polyhedra. The O-sublattice with ABABAB *hcp* stacking in $Al_2O_3$ (green zigzag lines) transforms into ABCABC *ccp* stacking in $MgAl_2O_4$ (blue straight lines), with a shared O plane at the interface (marked by red arrows). The emergence of structure type-A or -B depends on the position of the shared O layer at stacking A or B. Both structures have the same topotaxial OR between $Al_2O_3$ and $MgAl_2O_4$ as indicated by the coordination sketch in (c). (d) Z-contrast image of a larger area shows interface structures oscillating between type-A and type-B.

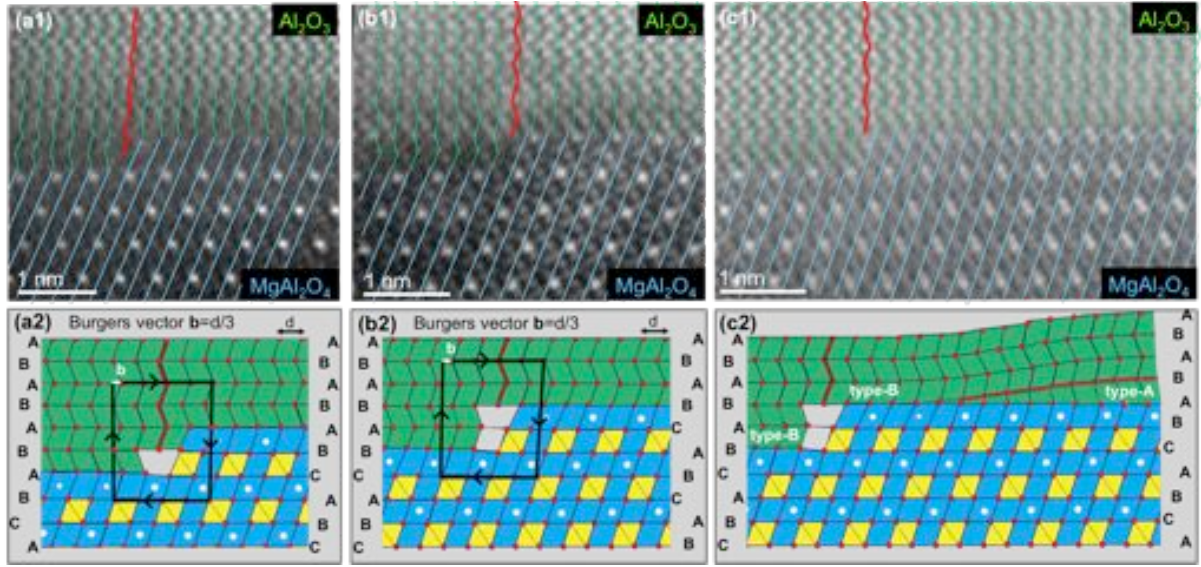

**Fig. 4**. (a1) and (b1) are STEM Z-contrast images showing partial dislocations associated with steps at $Al_2O_3$/$MgAl_2O_4$ interfaces for the type-A and type-B structures, respectively. The corresponding polyhedral models are shown in (a2) and (b2). The STEM Z-contrast image (c1) is a higher-magnification image of the step on the right side of Fig. 3d, showing a step-associated partial dislocation next to an interface structure type B-A transition. The corresponding polyhedral model is shown in (c2). The green zigzag lines and blue straight lines overlain on the Z-contrast images link O atoms between stacking planes, indicating the *hcp* and *ccp* O stacking in $Al_2O_3$ and $MgAl_2O_4$. The red zigzag lines show extra O ($\bar{2}110$) planes in $Al_2O_3$, which are terminated by misfit dislocations at the $Al_2O_3$/$MgAl_2O_4$ interface. The nearly horizontal red line in (c2) shows an extra O (0001) plane in $Al_2O_3$. The Burgers circuits are indicated by black arrows surrounding the dislocation cores in (a2) and (b2), resulting in a same Burgers vector modulus **b**=d/3. Burgers vectors "**b**" are shown by the white arrows to close the Burgers circuits. "d" corresponds to the O-O lattice spacing along $[11\bar{2}]_{MgAl2O4}$. The blue and green rhombs in the polyhedra models indicate Al at O octahedral sites in $Al_2O_3$ and $MgAl_2O_4$, while the yellow triangles indicate Mg at O tetrahedral sites in $MgAl_2O_4$. The white dots in the center of the blue rhombs indicate the "bright" Al columns at the pure Al-layers in $MgAl_2O_4$.

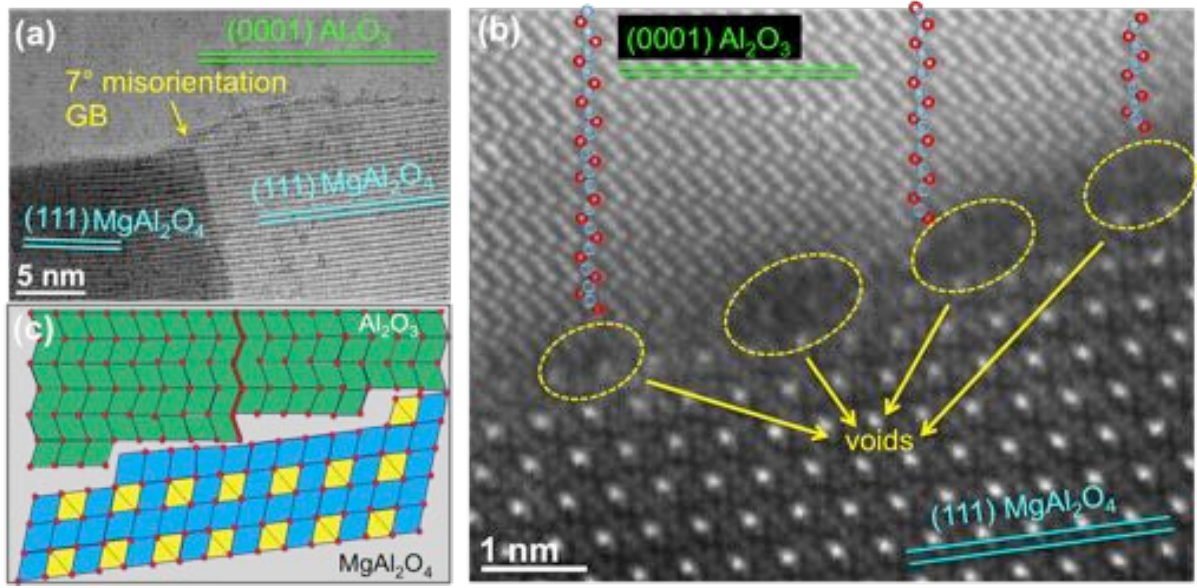

**Fig. 5.** Inclined $Al_2O_3$/$MgAl_2O_4$ interface for some $MgAl_2O_4$ nano-grains in the early growth stage. (a) STEM-BF image showing interface section where two $MgAl_2O_4$ nano-grains have 7° mutual misorientation around [1$\bar{1}$0] axis. (b) Z-contrast image showing the inclined interface section. Extra $<\bar{2}110>_{Al2O3}$ planes have been marked out with blue-red-blue-red circles showing zigzag Al-O-Al-O columns. (c) Illustration corresponding to the structure shown in (b). Green and blue rhombs indicate Al at O octahedral sites in $Al_2O_3$ and $MgAl_2O_4$, respectively, while the yellow triangles indicate Mg at O tetrahedral sites in $MgAl_2O_4$. Red zigzag line shows an extra O plane in $Al_2O_3$.

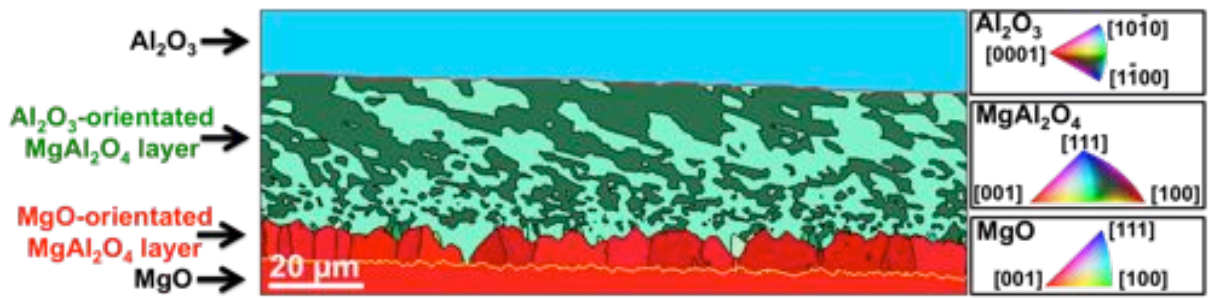

**Fig. 6.** EBSD IPF map of reconstructed grains showing $Al_2O_3$ (blue), MgO (red) and two layers of $MgAl_2O_4$: $Al_2O_3$-orientated $MgAl_2O_4$ including two twin orientations (dark and light green) and MgO-orientated $MgAl_2O_4$ (red). Boundaries with >2° misorientation angle are shown as black lines. The $MgAl_2O_4$/MgO interface is marked by a yellow line, the $Al_2O_3$/$MgAl_2O_4$ interface is marked by a red line. Adapted from (Li, Griffiths *et al.*, 2016) with permission.

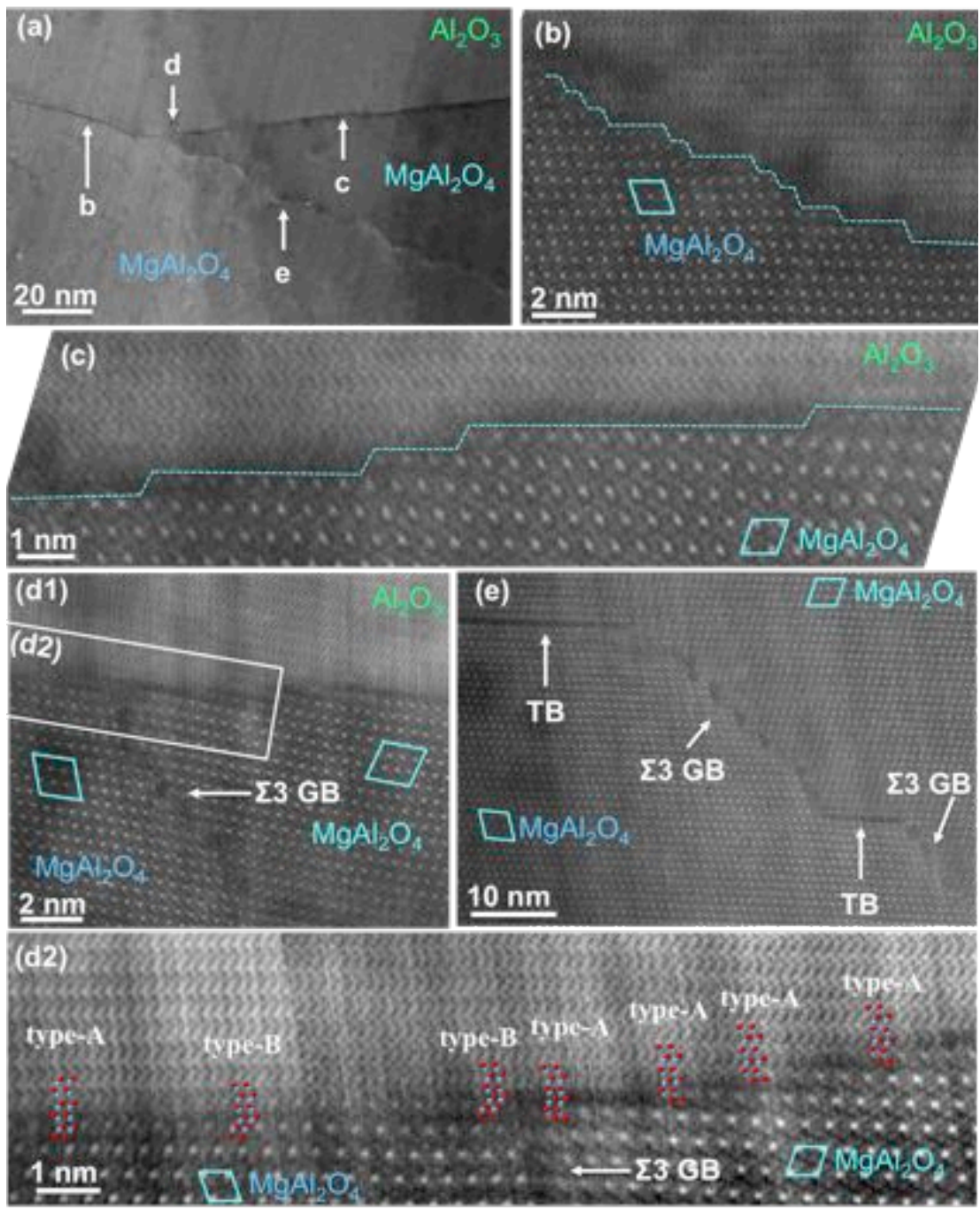

**Fig. 7.** Structure at MgAl$_2$O$_4$-MgAl$_2$O$_4$-Al$_2$O$_3$ triple-junctions. (a) Low-mag STEM BF image showing an Al$_2$O$_3$/MgAl$_2$O$_4$ interface that exhibits a sharp change of the Al$_2$O$_3$ facets on interface at a triple junction with a Σ3 MgAl$_2$O$_4$ GB. (b-e) High magnification STEM Z-contrast images recorded from the areas "b-e" marked in (a). The Al$_2$O$_3$/MgAl$_2$O$_4$ interface segments on the left (b) and the right (c) of the triple junction are both comprised of steps, which are facing in opposite directions. (d1-2) The interface structure at the triple junction. (d2) is a higher magnification image from area (d2) marked in image (d1), showing that directly at the triple junction the type-B interface on the left meets the type-A interface on the right. (e) The boundary between two MgAl$_2$O$_4$ grains is formed by incoherent Σ3 GBs and TBs. The TBs are parallel to the [111]$_{MgAl2O4}$ planes.

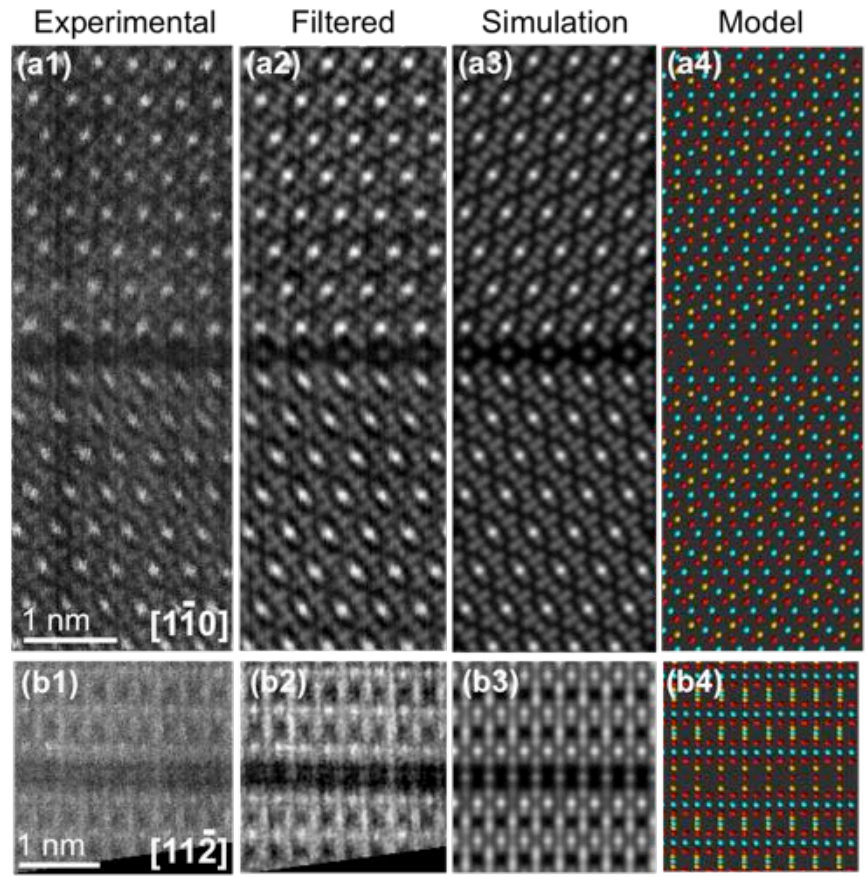

**Fig. 8.** Atomic structure of TB in $MgAl_2O_4$. (a) and (b) show the atomic configuration of a TB viewed along the $[1\bar{1}0]$ and the $[11\bar{2}]$ zone axes, respectively. (1), (2), (3) and (4) refer to experimental, filtered experimental, simulated Z-contrast images and atomic models. Blue, yellow and red circles in the models correspond to Al, Mg and O atoms, respectively.

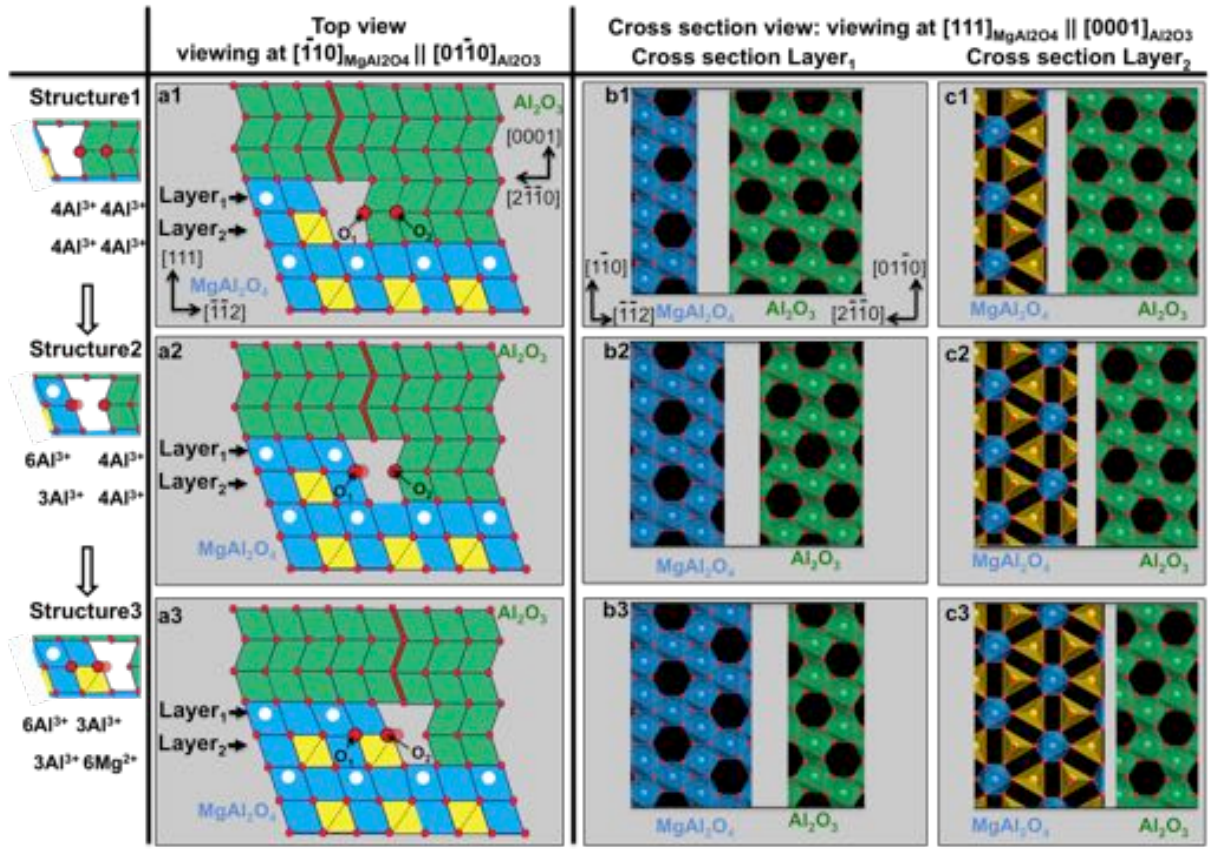

**Fig. 9.** Schematic illustration of the two elementary moves in the glide of a partial dislocation and associated migrations of a step in the $Al_2O_3/MgAl_2O_4$ interface. The interface step has a height of two close-packed O layers in $Al_2O_3$, and its migration is completed by the glide of the associated partial dislocation together with the simultaneous transformation of one Al-O layer of $Al_2O_3$ into an Al-O layer of $MgAl_2O_4$ (layer$_1$) and another Al-O layer of $Al_2O_3$ into an Al-Mg-O layer in $MgAl_2O_4$ (layer$_2$). The first elementary move in the glide process leads from structure 1 to structure 2, and the second move leads from structure 2 to structure 3, as shown in the leftmost panel of the illustration. Correspondingly, images a1-a3 show the interface models projected along the [110] axis of the $MgAl_2O_4$. b1-b3 and c-c3 show the top views of the (111)-parallel layer$_1$ and layer$_2$, respectively. The green color indicates the $AlO_6$ octahedrons in $Al_2O_3$; blue and yellow colors indicate the $AlO_6$ octahedrons and $MgO_4$ tetrahedrons in $MgAl_2O_4$, respectively. The half-transparent dashed red circle and the solid red circle in (a2) indicate the positions of column $O_1$ before and after glide, respectively.

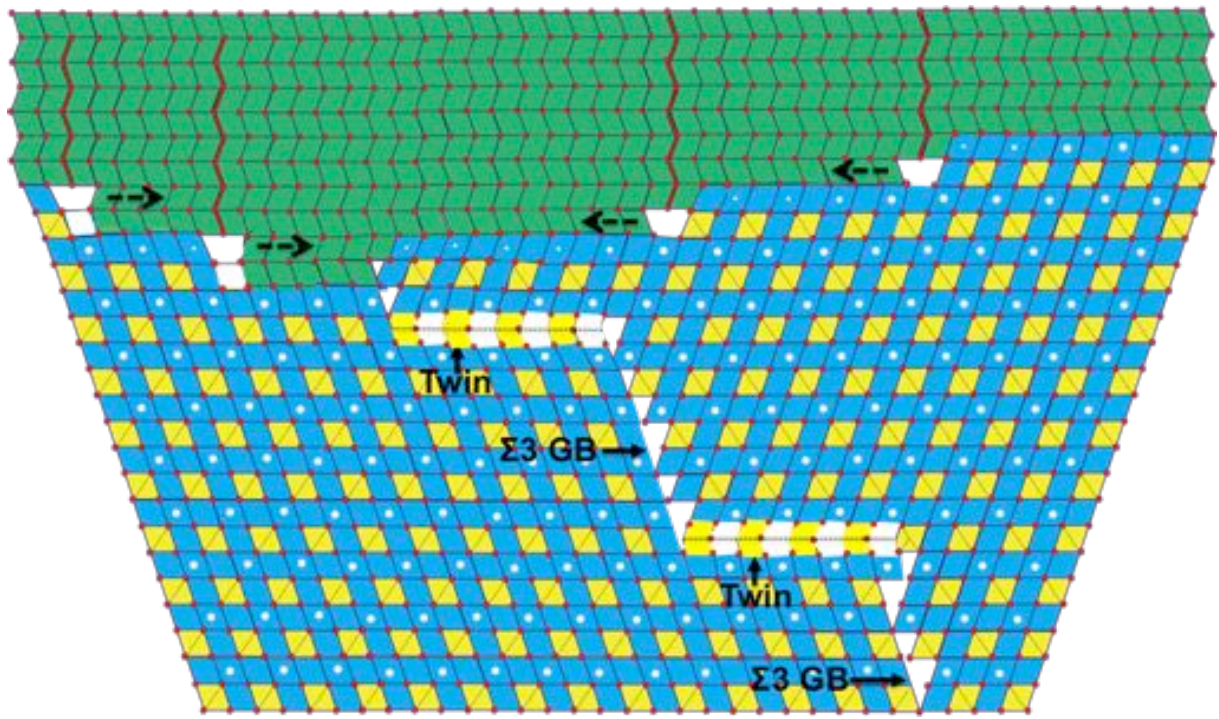

**Fig. 10.** Sketch showing the formation of TBs and Σ3 GBs in MgAl$_2$O$_4$ at the late grown stage.